\documentclass[twocolumn,showpacs,preprintnumbers]{revtex4}
\def\be{\begin{equation}}
\def\ee{\end{equation}}
\def\beq{\begin{eqnarray}}
\def\eeq{\end{eqnarray}}
\def\n{\nonumber}
\def\bay{\begin{array}}
\def\eay{\end{array}}

\begin{document}

\preprint{CIRI/02-smw03}
\title{Mach's Principle and Spatial Scale-Invariance of Gravity}
% Force line breaks with \\

\author{Sanjay M. Wagh}
\affiliation{Central India Research
Institute, Post Box 606, Laxminagar, Nagpur 440 022, India\\
E-mail:ciri@vsnl.com}

% \altaffiliation[Also at ]{School of Mathematical Sciences}
\bigskip

\date{February 4, 2002}
\bigskip

\begin{abstract}
Gravity does not provide any scale for matter properties. We argue
that this is also the implication of Mach's hypothesis of the
relativity of inertia. The most general spacetime compatible with
this property of gravity is that admitting three, independent
spatial homothetic Killing vectors generating an arbitrary
function of each one of the three spatial coordinates. The matter
properties for such a spacetime are (spatially) arbitrary and the
matter generating the spacetime admits {\it any\/} equation of
state. This is also the most general spacetime containing the weak
gravity physics in its entirety. This spacetime is machian in that
it is {\em globally\/} degenerate for anti-machian situations such
as vacuum, a single matter particle etc.\ and, hence, has no
meaning in the absence of matter.
\\
\centerline{Submitted to: Physical Review D - Brief Reports}
\end{abstract}

%\keywords{Suggested keywords}%
\maketitle

\newpage
Mach's principle is the hypothesis of the relativity of inertia
\cite{E39}. In a machian theory, the inertia of a body gets
determined by the presence of all other bodies in the universe.

It is well-known that the field equations of general relativity do
not incorporate Mach's principle. The point is that the field
equations admit curved vacuum solutions and also solutions that
are asymptotically flat. However, such solutions are anti-machian
since the existence of such solutions implies, in Einstein's own
words, that {\em the inertia [of a body] is {\/\em influenced\/}
by matter (at finite distances) but {\/\em not determined\/} by
it. If only a single mass point existed it would have inertia
...[but] in a consistent relativity theory there cannot be inertia
relative to ``space'' but only inertia of masses relative to each
other} \cite{subtle}. Therefore, the field equations of General
Relativity, expressed as only the {\/\em formal\/} equality of the
Einstein and the energy-momentum tensors, violate the machian
hypothesis of the relativity of inertia.

In this connection, it is important to stress that it is only the
equivalence principle which leads to the geometrization of
gravity. (It is on the basis of this principle that we demand the
local flatness of a spacetime manifold.) The equivalence principle
then leads, through a variational principle, to the field
equations of General Relativity. Hence, any theory of gravity
based on only the equivalence principle cannot be a machian
theory. This is certainly evident. Therefore, although Einstein
recognized \cite{e42} Mach's hypothesis of the relativity of
inertia as one of the basic principles in 1918, his enthusiasm for
it gradually waned and vanished soon afterwards \cite{schlipp}.

The overall influence of Mach's philosophy in the development of
general relativity is, however, undeniable. Mach's principle of
the relativity of inertia is also an important point of departure
from the newtonian thoughts. It, then, seems that we require,
perhaps, a radically new approach to reconcile Mach's ideas on the
relativity of inertia with those of the general covariance
\cite{fhjvn}.

We know that, speaking only mathematically, there can be many
four-dimensional spacetime geometries possible. From among these
possible geometries, the equivalence principle identifies those
that are locally Lorentzian. In a similar way, we need a principle
(other than the equivalence principle) that could be translated
into mathematical terms to identify {\em uniquely\/} the spacetime
geometries that are in conformity with the machian hypothesis. We
may then identify a {\em unique\/} set of equations for such
machian spacetimes and call them the ``correct'' field equations
of the machian theory of gravity. The ``correct'' field equations
must further obey the principle of general covariance that is a
fundamental symmetry of the classical Lagrangian formulation of
General Relativity.

Then, we may demand that, in a spacetime consistent with Mach's
principle, the metric tensor, and its all diffeomorphic forms,
should determine the inertial action and the metric tensor, and
its all diffeomorphic forms, should, in turn, be {\em
completely\/} determined by the mass distribution in the universe.
That is to say, the ``correct'' spacetimes, not as solutions of
the Einstein field equations but as solutions of the {\em new\/}
set of equations, should not be obtainable in the absence of
matter. (To quote Pais \cite{subtle}, ``Einstein never said so
explicitly, but it seems reasonable to assume that he had in mind
that the correct equations should have no solutions at all in the
absence of matter.'')

Therefore, although the Einstein field equations are not machian,
there could be another set of equations leading us to spacetime
geometries of the sort implied by Mach's hypothesis.

The quest for such a set of equations on the basis of Mach's
version of the hypothesis has, however, not led us to any definite
conclusions in the past. It has been difficult to identify
spacetime geometries uniquely from the statement of Mach's
principle since it is difficult to translate it into some precise
mathematical language.

There could, however, be an implication of Mach's principle which
is translatable to a precise mathematical requirement to be
imposed on the spacetime apart from that of its local flatness. It
is in this spirit that the present paper shows that the Einstein
field equations augmented with the principle of spatial scale
invariance lead to manifestly machian relativity theory of
gravity.

The point of departure from the usual analysis dealing with Mach's
principle is the following observation: Mach's principle states
that the inertia of a particle of matter is the result of its
interaction with all other particles in the universe.
Consequently, there must be energy density of matter
``everywhere'' in a machian universe. This can be interpreted to
mean that we can assemble ``masses'' to produce another ``mass''
and that the process of this building up of mass cannot be
terminated in space. This is, then, recognized as the principle of
the mass-scale and/or spatial scale invariance of the theory of
gravity. Therefore, we note an {\em important aspect of Mach's
hypothesis of the relativity of inertia - that the spatial
scale-independence of gravity is (one of) its direct
implications}.

To further fix our ideas, we first note that the phenomenon of
gravitation does not provide any length-scale or mass-scale for
spatial distributions of matter properties. The scale-independence
of Newtonian gravity applies only to space and not to time.
Newton's law of gravitation does not specify any property of
matter that it deals with. It applies irrespective of the form of
matter under consideration. To be precise, we can assemble masses
to produce another mass, of any desired spatial density
distribution as well as of any size and made up of any material.
Newtonian law of gravitation permits this even when other physical
phenomena are considered together with that of gravitation. The
spatial scale-independence and/or mass-independence are important
properties of the newtonian gravity.

Then, accordingly, the newtonian theory is also a machian theory
but it is not a complete theory of relativity since it is based
only on the galilean invariance. A theory of gravity that is based
on the principle of general covariance is the theory of General
Relativity.

In General Relativity, the spatial scale and the mass scale become
one since its fundamental constants provide us the means to
convert length to mass and vice-versa. Therefore, if the spatial
scale-independence is any basic property of gravity then, General
Relativity must admit a spacetime with matter density as an
arbitrary function of {\em each\/} of the three spatial
coordinates, in general. We emphasize that such a spacetime metric
and all other metric forms that are reducible to it under
non-singular coordinate transformations, that is to say,
diffeomorphic to it, are the only solutions of the field equations
of General Relativity that are consistent with gravity not
possessing a length-scale for matter properties.

In this connection, we note that the field equations of General
Relativity were arrived at by demanding only that these reduce to
the Newton-Poisson equation in the weak gravity limit
\cite{eg1913, subtle}. But, the {\em field equations of any theory
of gravity should contain the entire weak gravity physics due to
the applicability of the laws of weak gravity to any form of
matter displaying any physical phenomena}. The field equations are
expected to be only the {\em formal\/} equality of the appropriate
tensor from the geometry and the energy-momentum tensor of matter.
Then, the field equations of General Relativity could have been
obtained by imposing the requirement that these reduce to the
single ``equation of the entire weak gravity physics''.

However, there is no ``single'' equation for the ``entire weak
gravity physics" since we include different physical effects in an
ad-hoc manner in the newtonian physics.

But, {\em there can be a ``single'' spacetime containing the
entire weak gravity physics}. Therefore, we need a principle to
identify such a solution of the field equations. In the weak field
limit, the spatial scale-invariance is the freedom of
specification of matter properties through three independent
functions of the three spatial coordinates, in general.

{\em The spatial scale-invariance is then the principle that could
help us identify spacetimes containing the entire weak gravity
physics.} Indeed. the spatial scale invariance identifies
(\ref{genhsp}) as the single such spacetime \cite{cch}. It has
appropriate energy-momentum fluxes, applicability to any form of
matter and, hence, it contains the entire weak gravity physics.
The newtonian law of gravitation then gets replaced by the single
general relativistic spacetime of (\ref{genhsp}) that contains all
of the weak gravity physics. But, spatial scale-independence needs
to be separately imposed on the field equations to obtain it.

In what follows, we recall from \cite{cch} the most general
spatially scale-invariant spacetime for the sake of completeness
here.

In general, a homothetic Killing vector captures \cite{carrcoley}
the notion of the scale-invariance. A spacetime that conforms to
the spatial scale-invariance, to be called a {\em spatially
homothetic spacetime}, is then required to admit an appropriate
{\em spatial\/} homothetic Killing vector ${\bf X}$ satisfying \be
{\cal L}_{\bf X} g_{ab}\;=\;2\,\Phi\,g_{ab} \label{hkveq} \ee
where $\Phi$ is an arbitrary constant.

We then expect spatially homothetic spacetimes to possess
arbitrary spatial characteristics for matter. This is also the
broadest (Lie) sense of the scale-invariance of the spacetime
leading not only to the reduction of the Einstein field equations
as partial differential equations to ordinary differential
equations but leading also to their separation. That is why we
expect to obtain three arbitrary functions for a spacetime
admitting three independent spatial homothetic Killing vectors.

In general, we then demand that the spacetime admitting no special
symmetries, that is no proper Killing vectors, admits {\em
three\/} independent homothetic Killing vectors corresponding to
the three dimensions for which gravity provides no length-scale
for matter inhomogeneities. Such a metric, from the broadest (Lie)
sense, admits three functions $X(x)$, $Y(y)$, $Z(z)$ of three
space variables, conveniently called here, $x$, $y$, $z$, each
being a function of only one variable.

Based on the above considerations, we then demand that there exist
three independent spatial homothetic Killing vectors \beq {\bf
{\cal X}} = (0, f(x), 0, 0) \label{genhkv1} \\ {\bf {\cal Y}} =
(0, 0, g(y), 0) \label{genhkv2} \\ {\bf {\cal Z}} = (0, 0, 0,
h(z)) \label{genhkv3} \eeq for the general spacetime metric \be
ds^2\;=\;g_{ab}dx^adx^b \ee with $g_{ab}$ being functions of the
coordinates $(t, x, y, z)$. Here each of the three vectors ${\cal
X}$, ${\cal Y}$ and ${\cal Z}$ satisfies (\ref{hkveq}) with
$\Phi_x$, $\Phi_y$, $\Phi_z$ as corresponding constants.

Then, the spacetime metric is, {\em uniquely\/}, the following
\begin{widetext} \beq ds^2 = &-&\,X^2(x)\,Y^2(y)\,Z^2(z)\,dt^2
\,+\,\gamma_x^2\left(\,\frac{dX}{dx}\,\right)^2Y^2(y)\,Z^2(z)\,
A^2(t) \,dx^2 \n \\&+&
\gamma_y^2\,X^2(x)\left(\,\frac{dY}{dy}\,\right)^2
Z^2(z)\,B^2(t)\,dy^2\,+\,
\gamma_z^2\,X^2(x)\,Y^2(y)\left(\,\frac{dZ}{dz}\,\right)^2C^2(t)\,dx^2
\label{genhsp} \eeq
\end{widetext} where $\gamma$ s are constants related to $\Phi$ s.
This is the most general spacetime compatible with gravity not
possessing any length-scale for matter inhomogeneities in its {\em
diagonal\/} form.

The coordinates $(t, x, y, z)$ in which this metric is separable
are co-moving. Hence, the matter 4-velocity is $U^a\,=\,(U^t, U^x,
U^y, U^z)$ with all the components non-vanishing in general. Then,
using, for example, the software {$\scriptstyle{\rm SHEEP}$}, it
is easy to see that the Einstein tensor has appropriate components
\begin{widetext} \beq G_{tt}&=&
-\frac{1}{\gamma_x^2A^2}-\frac{1}{\gamma_y^2B^2}
-\frac{1}{\gamma_z^2C^2} + \frac{\dot{A}\dot{B}}{AB} +
\frac{\dot{A}\dot{C}}{AC} +\frac{\dot{B}\dot{C}}{BC}\\
G_{xx}&=&\gamma_x^2A^2 \left(\frac{L_{,x}}{L}\right)^2
\left[-\frac{\ddot{B}}{B}-\frac{\ddot{C}}{C} -
\frac{\dot{B}\dot{C}}{BC}+\frac{3}{\gamma_x^2A^2}+
\frac{1}{\gamma_y^2B^2} + \frac{1}{\gamma_z^2C^2}\right]
\\G_{yy}&=&\gamma_y^2B^2 \left(\frac{M_{,y}}{M}\right)^2
\left[-\frac{\ddot{A}}{A}-\frac{\ddot{C}}{C} -
\frac{\dot{A}\dot{C}}{AC}+\frac{3}{\gamma_y^2B^2}+
\frac{1}{\gamma_x^2A^2} + \frac{1}{\gamma_z^2C^2}\right]
\\G_{zz}&=& \gamma_z^2C^2 \left(\frac{N_{,z}}{N}\right)^2
\left[-\frac{\ddot{A}}{A}-\frac{\ddot{B}}{B} -
\frac{\dot{A}\dot{B}}{AB}+\frac{3}{\gamma_z^2C^2}+
\frac{1}{\gamma_x^2A^2} + \frac{1}{\gamma_y^2B^2}\right] \\
G_{tx}&=&2\frac{\dot{A}L_{,x}}{AL} \\
G_{ty}&=&2\frac{\dot{B}M_{,y}}{BM}\\
G_{tz}&=&2\frac{\dot{C}N_{,z}}{CN} \\
G_{xy}&=&2\frac{L_{,x}M_{,y}}{LM} \\
G_{xz}&=&2\frac{L_{,x}N_{,z}}{LN} \\
G_{yz}&=&2\frac{M_{,y}N_{,z}}{MN} \eeq \end{widetext}
corresponding to expected non-vanishing energy-momentum fluxes.

In general, matter distribution in the weak gravity physics has a
newtonian gravitational force along the radial vector joining the
chosen location of a given matter particle to the origin of the
``cartesian'' coordinate system. Consequently, when this force is
unbalanced, we expect non-vanishing energy-momentum fluxes along
all the coordinate directions. Therefore, the energy-momentum
tensor of matter in the spacetime of (\ref{genhsp}) is, in
general, imperfect and/or anisotropic.

Therefore, matter in the spacetime could, in general, be {\em
imperfect\/} or {\em anisotropic\/} indicating that its
energy-momentum tensor could be
\begin{widetext} \beq {}^{\rm
I}T_{ab}&=&(\,p\,+\,\rho\,)\,U_a\,U_b \;+\; p\, g_{ab}
\;+\;q_a\,U_b \;+\; q_b\,U_a \;-\;2\,\eta\,\sigma_{ab} \\ {}^{\rm
A}T_{ab}&=&\rho\, U_a\,U_b \;+\; p_{||}\,n_a\,n_b \;+\;
p_{\bot}\,P_{ab} \eeq
\end{widetext} where $U^a$ is the matter 4-velocity, $q^a$ is the
heat-flux 4-vector relative to $U^a$, $\eta$ is the
shear-viscosity coefficient, $\sigma_{ab}$ is the shear tensor,
$n^a$ is a unit spacelike 4-vector orthogonal to $U^a$, $P_{ab}$
is the projection tensor onto the two-plane orthogonal to $U^a$
and $n^a$, $p_{||}$ denotes pressure parallel to and $p_{\bot}$
denotes pressure perpendicular to $n^a$. Also, $p$ is the
isotropic pressure and $\rho$ is the energy density. Note that the
shear tensor is trace-free. We will represent by $\sigma$ the
shear-scalar that is given by $\sqrt{6}\;\sigma$.

The energy momentum tensor of matter essentially describes the
physical effects due to all other reasons apart from the
gravitational ones. The gravitational effects are determined by
the spacetime geometry that is given by (\ref{genhsp}). That the
principle of spatial scale-invariance uniquely determines the
spacetime geometry irrespective of the form of the energy-momentum
tensor of matter is therefore clear.

It is then easy to see that the field equations do not determine
the spatial functions $X(x)$, $Y(y)$, $Z(z)$. Further, the density
is initially non-singular for nowhere-vanishing spatial functions
$X(x)$, $Y(y)$, $Z(z)$. Moreover, the temporal functions $A(t)$,
$B(t)$, $C(t)$ get determined only from the properties of matter
generating the spacetime, such as its equation of state, shear
tensor etc.

Further, the spacetimes admitting homothetic Killing vectors of
the form \be (T, \bar{x}, \bar{y}, \bar{z}) \label{usualgenhkv}
\ee or, combinations thereof, are contained with (\ref{genhsp})
provided the transformations of (\ref{genhkv1}) - (\ref{genhkv3})
leading to (\ref{usualgenhkv}) are non-singular. In other words,
the spacetime of (\ref{genhsp}) will be ``diffeomorphic'' to some
of the spacetimes admitting (\ref{usualgenhkv}).

Further, we note that symmetries can, in some appropriate sense,
be {\em locally\/} imposed on the spacetime of (\ref{genhsp}) so
that we obtain {\em locally\/} symmetric objects in it. This will,
however, require a detailed analysis of the metric (\ref{genhsp})
than presented in this short exposition.

Before concluding, we emphasize that the metric (\ref{genhsp})
becomes singular for {\em globally\/} uniform density which
requires $X(x)\,=\,{\rm constant}$ for all $x$ and $Y(y)\,=\,{\rm
constant}$ for all $y$ and $Z(z)\,=\,{\rm constant}$ for all $z$.
It also follows that the metric (\ref{genhsp}) is singular for
global vacuum solutions which require $X(x)\,=\,\infty$ for all
$x$ and $Y(y)\,=\,\infty$ for all $y$ and $Z(z)\,=\,\infty$ for
all $z$. Hence, the metric (\ref{genhsp}) cannot be a uniform
density or a single particle solution to the field equations.
Further, it is also clear that the metric (\ref{genhsp}) cannot be
asymptotically flat without any such conditions on the metric
functions that make it singular there.

Then, we note here that the spacetime of (\ref{genhsp}) is also
the most general, spatially homothetic, cosmological spacetime. It
may further be emphasized that the spacetime of (\ref{genhsp}) is
Machian in the sense that it becomes {\em globally\/} singular in
evidently anti-machian situations such as a single particle of
matter, vacuum etc. Therefore, Mach's hypothesis of the relativity
of inertia, interpreted to mean the spatial scale-invariance of
the theory of gravity, does identify a unique solution of the
Einstein field equations satisfying it.

In his Autobiographical Notes, Einstein remarked \cite{schlipp}:
``{\it Mach conjectures that in a truly rational theory inertia
would have to depend upon the interaction of the masses, precisely
as was true for Newton's other forces, a conception which for a
long time I considered as in principle the correct one. It
presupposes implicitly, however, that the basic theory should be
of the general type of Newton's mechanics: masses and their
interaction as the original concepts. The attempt at such a
solution does not fit into a consistent field theory, as will be
immediately recognized.}''

However, we have essentially shown here that Mach's hypothesis of
the relativity of inertia does possess a consistent field
theoretical formulation when we recognize the spatial scale
invariance of gravity as one of its direct implications.
%\newpage
\section*{Acknowledgements}
Dedication: This article is dedicated to the memory of Sir Fred
Hoyle. His courage in propounding new ideas, his enthusiasm for
discussing various alternative ideas and his deep, almost
mysterious, insights into difficult physical problems will be
missed by all who ever interacted with him, no matter for short or
long durations.

I am grateful to Kesh Govinder for stimulating discussions as well
as for useful collaborations and to Malcolm MacCallum for
providing me the software $\scriptstyle{\rm SHEEP}$ that has been
used to perform the calculations presented here.

\end{document}